\documentclass[a4paper]{article}
\usepackage[left=2.6cm,right=2.6cm,top=3.3cm,bottom=4.2cm]{geometry}
\usepackage{apalike}
\usepackage[francais]{babel}
\usepackage[T1]{fontenc}
\usepackage[utf8]{inputenc}
\usepackage{csquotes}
\usepackage{epsfig}
\usepackage{subfigure}
\usepackage{calc}
\usepackage{amstext}
\usepackage{amsmath}
\usepackage{amsthm}
\usepackage{multicol}
\usepackage{pslatex}
\usepackage{algorithm,algorithmicx,algpseudocode,SCITEPRESS}
\usepackage{subfigure}
\usepackage{tikz}
\usepackage{amssymb, bm}
\usepackage{mathtools, bm}
\usepackage{apalike}
\usepackage{caption}
\usepackage{url}
\usetikzlibrary{arrows,automata}

\numberwithin{equation}{section}

\def\spr#1{{\rm range}(#1)}
\def\pref#1{{\rm Pref}(#1)}

\def\sprt#1#2{ {\rm sprt}(#1,#2)}

\def\P#1{\mathop{\mathbb{P}_{#1}}}
\def\Sr{\mathop{\mathbb{S}}}
\def\S#1{\mathop{\mathbb{S}_{#1}}}

\def\F{\mathop{\mathbb{F}}}
\def\T{\mathop{\mathbb{T}}}

\def\AP#1{{\cal P}_{#1}}
\def\AS#1{{\cal S}_{#1}}

\def\b#1{\overline{#1}}

\newcommand{\B}{\mathbb B}
\newcommand{\K}{\mathbb K}
\newcommand{\M}{\mathbb M}
\newcommand{\N}{\mathbb N}

\renewcommand{\thefigure}{\arabic{figure}}

\newtheorem{theorem}{Theorem}[section]
\newtheorem{lemma}[theorem]{Lemma}
\newtheorem{proposition}[theorem]{Proposition}
\newtheorem{corollary}[theorem]{Corollary}
\newtheorem{definition}[theorem]{Definition}

\newenvironment{proof}[1][Proof]{\begin{trivlist}
		\item[\hskip \labelsep {\bfseries #1}]}{\end{trivlist}}

\begin{document}

\title{Mining Frequent Itemsets	a Formal Unification\thanks{This work is supported by the MESRS-Algeria under the project number 8/U03/7015}}

\author{\authorname{Slimane Oulad-Naoui\sup{1}, Hadda Cherroun\sup{2} and Djelloul Ziadi\sup{3}}
	\affiliation{\sup{1}Département des Mathématiques et d'Informatique, Université de Ghardaia, Ghardaia, Algeria}
	\affiliation{\sup{2}Laboratoire d'Informatique et de Mathématiques, Université Amar Telidji, Laghouat, Algeria}
	\affiliation{\sup{3}Laboratoire LITIS - EA 4108, Normandie Université, Rouen, France}
	\email{s.ouladnaoui@univ-ghardaia.dz, hadda\_cherroun@mail.lagh-univ.dz, djelloul.ziadi@univ-rouen.fr}}


\abstract{It is generally well agreed that developing a unifying theory is one of the most important issues in Data Mining research. In the last two decades, a great deal of work has been devoted to the algorithmic aspects of the Frequent Itemset (FI) Mining problem. We are motivated by the need for formal modeling in the field. Thus, we introduce and analyze, in this theoretical study, a new model for the FI mining task. Indeed, we encode the itemsets as words over an ordered alphabet, and state this problem by a formal series over the counting semiring $(\N,+,\times,0,1)$, whose range constitutes the itemsets and the coefficients are their supports. This formalism offers many advantages in both fundamental and practical aspects: the introduction of a clear and unified theoretical framework through which we can express the main FI-approaches, the possibility of their generalization to mine other more complex objects, and their incrementalisation or parallelisation; in practice, we explain how this problem can be seen as that of word recognition by an automaton, allowing an efficient implementation in $O(|Q|)$ space and $O(|\mathcal{F}_L||Q|])$ time, where $Q$ is the set of states of the automaton used for representing the data, and $\mathcal{F}_L$ the set of prefixial longest FI.}

\maketitle
\section{\uppercase{Introduction}}
\noindent Mining Frequent Itemsets (FI) is an important problem in Data Mining (DM). Although primitive, it constitutes one of the most challenging and over-two-decade-well-studied subject in the field. Since the introduction of the Apriori algorithm by Agrawal~\cite{Agrawal94}, several algorithms have been proposed to solve it. Without claim of exhaustiveness, we can categorize these works into three main classes, for more detail see~\cite{Hipp00,fimi03,Han07}: (\textit{i}) Enumeration of all FI, (\textit{ii}) Discovery of closed/maximal FI, and (\textit{iii}) Incremental algorithms.\\
The first class of algorithms aims to extract the whole set of FI. The problem space exploration approaches used can be distinguished by the traversal and the support calculation methods~\cite{Hipp00}. In level-wise techniques, a breadth-first traversal is adopted, where a $k$-itemset is derived by extending a frequent one of length $k-1$~\cite{Agrawal94}. The calculation of the support of an itemset is performed by database scans. In these techniques, the most remarkable is, undeniably, the A-priori heuristic used to prune the problem space, and widely used in all algorithms later. Unfortunately, these techniques suffer from two major drawbacks: Generating a huge number of candidates, and an excessive I/O cost needed for support counting.\\
In~\cite{Zaki00}, the author considered the data from a vertical angle of view, where he associated with each itemset $X$ its list of transactions (tidlist) and uses set intersection for support calculation, which has proven to be a more effective trick. However, and despite the common-prefix equivalence relation proposed to decompose the problem, this approach requires, in dense datasets particularly, a large time and intermediate memory to perform intersections.\\
In order to reduce the size of the dataset and avoid multiple scans of it, Han et al. introduced the FPGrowth algorithm~\cite{Han04} that uses a compact structure called Frequent-Pattern Tree (FPTree) enhanced with the itemset supports. This algorithm generates recursively a grown-pattern conditional database projections for which a corresponding FPTrees are also constructed. Though, the performance gain shown, the original version of FPGrowth induces, sadly, an abundant memory and time overhead due to repetitive sorting and reconstructions.\\   
The algorithms of the second class focus on the minimum set of FI, called the cover, which allows to generate all the rest~\cite{Pasquier99}. Thereby, the closed and maximal FI notions have been introduced. These approaches use Formal Concept Analysis (FCA)~\cite{Wille82} to extract the set of frequent concepts, that constitutes a condensed representation of the entire set of FI.\\
The concern of the algorithms of the third class was the incrementality. That is, how to generate the set of FI, and to maintain it in the case of dynamic datasets~\cite{Valtchev08}. Here, the same philosophies were adopted in either algorithmic fashion or using FCA.\\
Summing up, after more than two decades of active research on the subject, with countless techniques including various efficient algorithms and judicious data structures each with its advantages and drawbacks, we believe that it will be convenient to go back and ask a key question: Besides the existing ones~\cite{Godin95,Zaki98}, are there other formalisms for this basic problem ? In other words, we aim to develop a general unifying model able to express the works done so far in the main state of the art approaches. We wish, moreover, that the proposed formalism should enjoy some capitals characteristics such as : the completeness while remaining simple and intuitive, extensibility, and efficiency. That is, provide, why not, an implementation having a better performances, if not stand at least comparable to those of the existing techniques.\\
The elaboration of unifying models is a well-established issue in DM~\cite{Yang06}. We postulate that the unification can be facilitated if we focus on a particular DM-task. In this paper, we address this question for the FI-mining problem. Indeed, we introduce a new model for enumerating all FI based on formal series, which meets the above proprieties. First, it defines a unified theoretical framework, which leads to see the equivalence of the algorithms as stipulated in~\cite{fimi03} and confirmed in one of the early comparative studies~\cite{Hipp00}. Second, it allows their generalization for mining more complex objects. We prove also a natural decomposition scheme, often required in many aspects of the problem such that incrementality or parallelization. Moreover, we explain how this problem can be transposed to that of the realization of a formal series by a weighted automaton~\cite{salomaa78}, and consequently, to that of word recognition, which is a largely invested topic with a very mature algorithmic. Finally, we propose an efficient algorithm to enumerate all FI, which runs in place without extra memory.\\   
The remaining of this paper is organized as follows. We begin, in Section 2, by some preliminaries on the basic concepts and notions to be used throughout this article. In Section 3, we recall the FI mining problem, and introduce our model. Section 4 is devoted to the definition, the proofs, the construction of the proposed automaton, and the analysis of the mining algorithm. In Section 5, we discuss our model against the existing techniques and show how these can be derived from it, and conclude in Section 6 with some extensions.

\section{\uppercase{Preliminaries}}
	\noindent A set $\M$  with an associative binary internal operation $\ast$ admitting a unique $e\in\M$ as an identity element forms a structure of \textit {monoid}, which we denote $(\M,\ast,e)$. When the operation $\ast$ is also commutative then the monoid is commutative. The popular example is the free monoid $A^*$ of the set of words over an alphabet $A$ equipped with the concatenation of two words, and having the empty word $\varepsilon$ as an identity element.\\
	A word $u$ is a \textit{prefix} (respectively a  \textit{suffix}) of a word $w$ if there exists a word $v$ such that $w=uv$ (respectively $w=vu$). The set of the prefixes of a word $u$ will be denoted $\pref{u}$. This concept can be extended to a set of words by performing the union of the prefixes of its elements. A word $u=u_1\ldots u_k$ is a \textit{subsequence} of a word $w=w_1\ldots w_l\; (k \le l )$ if there exist words $v_1 ,\ldots ,v_{k+1}$, such that $w=v_1u_1v_2u_2\ldots v_ku_kv_{k+1}$. We write then $u \preccurlyeq w$.\\
	A \textit{semiring} is a tuple $(\K,+,\times,0,1)$ such that: $(\K,+,0)$ is a commutative monoid, $(\K,\times,1)$ is a monoid, $\times$ distributes on both sides over $+$, and $0$ is an absorbing element with respect to $\times$. Examples of semirings are $(\N,+,\times,0,1)$ of positive integers, $(\B,\lor,\land,\bot,\top)$ of booleans, and the tropical semiring $(\N\cup\{\infty\}, min, +,\infty,0)$.\\
	Over the monoid $A^*$, we define a \textit{formal series} $\Sr$ with coefficients in a semiring $\K$ as a mapping~$\Sr:  A^*\rightarrow  \K$, which associates with each $w$ its \textit{coefficient} $\langle \Sr,w \rangle$. The series $\Sr$ itself will be written as a sum:	
	
	\begin{equation}
	\Sr=\sum_{w\in A^*} \langle \Sr,w \rangle w
	\end{equation}
	
	The set $ \spr{\Sr} =\{w \in A^* |\; \langle \Sr,w \rangle \neq 0 \} $ of words with non-null coefficients is called the \textit{range} of the series $\Sr$ (also called its support, but we prefer range to avoid the confusion with the support of an itemset).
	The set of formal series over $A$ with coefficients on $\K$ is denoted $\K\langle\langle A\rangle\rangle$. A structure of a semiring is defined on $\K\langle\langle A\rangle\rangle$ as follows, $\Sr$ and $\T$ are two formal series on $A$ with coefficients in $\K$:
	\begin{equation}
	\langle \Sr+\T,w \rangle =\langle \Sr,w \rangle + \langle \T,w \rangle
	\end{equation}
	\begin{equation}
	\langle \Sr\T,w \rangle=\sum_{ uv=w}\langle \Sr,u \rangle \langle \T,v \rangle
	\end{equation}
	The subset of the series of $\K \langle\langle A \rangle\rangle$ with finite range are called \textit{polynomials} and denoted by $\K \langle A\rangle$. Thereafter, in the case of the monoid $A^*$, and for its identity element $\varepsilon$,  and for all  $k\in\K$ we write $k\varepsilon$ (or simply $k$) the term having $k$ as coefficient of $\varepsilon$. In the same way, for a word $w$ on $A^*$, we denote $kw$ (respectively $w$) the term whose coefficient for $w$ is $k$ (respectively $1$).\\
	A \textit{weighted automaton} $\mathcal{A}$ over an alphabet $A$ with coefficients in a semiring $\K$ is a tuple $\mathcal{A} = (Q,A,\mu,\lambda,\gamma)$, where $Q$ is the finite set of states, $\mu$ the function from $Q$ to $\K$ of the initial (input) weights, $\lambda$ the function from $Q\times A\times Q$ to $\K$ of the transition weights, and $\gamma$ the mapping from $Q$ to $\K$ of the final (output) weights. A path $c$ in $\mathcal{A}$ is a succession of transitions:  $(q_0,a_1,q_1)\ldots(q_{n-1},a_n,q_n)$ labelled by the word $a_1\ldots a_n$ obtained by the concatenation of the symbols of its edges. Its weight is the product of the weights of its transitions:
	\begin{equation}
	\omega(c)=\mu(q_0)\lambda(q_0,a_1,q_1)\ldots \lambda(q_{n-1},a_n,q_n)\gamma(q_n)
	\end{equation}
	If we denote by $\mathcal{C}(u)$ the set of all paths labelled $u$. The weight of a word $u$ in the automaton $\mathcal{A}$, denoted $\mathcal{A}(u)$, is the sum of the weights of the elements of $\mathcal{C}(u)$:
	
	\begin{equation}
	\mathcal{A}(u)=\sum_{c\,\in\,\mathcal{C}(u)}\omega(c)
	\end{equation}
	The size of an automaton is the number of its transitions.
	\section{\uppercase{Problem Statement and Notations}}
	\noindent First, let us recall the basic concepts of the FI mining problem, and introduce the definitions and notations used throughout this paper.\\
	Let $A =\{a_1,a_2,\ldots,a_m\}$ be an alphabet of $m$ symbols called \textit{items}. Those can designate, according to the application domain, a products purchased from a supermarket, a visited Web pages, a collection of attributes$\ldots$etc. An \textit{itemset} is a subset of A, if $k$ is its cardinal it is called a $k$-itemset. A \textit{transaction} $t_i$ is a nonempty set of items identified by its unique identifier $i$. A \textit{dataset} $D$  is a set of $n$ transactions,  which we denote as a multi-set: $D =\{t_1,t_2,\ldots,t_n\}$. In a dataset $D$, the \textit{support} of an itemset $x$, denoted $\sprt{x}{D}$, is the number of transactions containing $x$, i.e :
	\begin{equation}
	\sprt{x}{D} = |\{t_k \in D \;\mid\; x \subseteq t_k \}|
	\end{equation}
	An itemset $ x$ is \textit{frequent} if its support exceeds a specified minimum support-threshold $s$. That is, $ x$ is frequent in $D$ if and only if $\sprt{x}{D} \geq s$. A frequent itemset is \textit{maximal} if an only if there is no superset of it which is frequent.    
	The problem of mining FI consists to discover the set $\mathcal{F}$ of all itemsets whose support is greater than the given minimum support-threshold $s$.
	\subsection{The Polynomial Model}
\noindent Now, we show how to translate the FI mining problem to the formal series model. Taking into account the finiteness of the modeled data (itemsets and datasets), we adopt thus a modeling based on polynomials.\\
The main idea in this modeling is to encode an itemset by a word, and all its subsets by a polynomial.  After defining the polynomial of a dataset, the question is then to extract from this polynomial all the terms where the support-criterion holds.\\
First, let us assume, without loss of generality, that the alphabet $A$ is sorted according to an arbitrary total order, where we can write:\\
 $A =\{a_1,a_2\ldots a_m \} \text{, with: }  \varepsilon < a_1 <  a_2 <\ldots < a_m$. We represent a $k$-itemset $x=\{a_ {i_1}, a_ {i_2}, \ldots, a_{i_k} \}$ by the word $w(x)$ of length $k$,  built by the concatenation of its items according to the predefined order. We will write: 
\begin{equation}
	w(x) =a_{i_1}a_{i_2}\ldots a_{i_k}, \text{ such that }a_{i_1}<\ldots <a_{i_k}
\end{equation}
In what follows, we confuse an itemset $x$ and its word representation $w(x)$. That is, instead of $x=\{a,b,c\}$, we write simply $x=abc$. Note that the empty itemset $\emptyset$ is represented by the empty word $\varepsilon$ of length zero ($|\epsilon| = 0$).
\begin{definition}[Itemset Subsequence Polynomial]
		Let $x =a_{i_1}a_{i_2}\ldots a_{i_k}$ be a $k$-itemset, The subsequence polynomial $\S{x}$ associated with $x$ is defined as follows:
		\begin{equation}
		\S{x}= (a_{i_1}+1)(a_{i_2}+1)\ldots(a_{i_k}+1) \text{, with:  }\S{\varepsilon} = 1
		\end{equation}
		\label{def1}
\end{definition}
Hereafter, we denote, for each $a\in A$, by $\b{a}$ the polynomial $(a+1)$. So, the polynomial $\S{x}$ associated with a $k$-itemset $x$ will be denoted: $\S{x}=\b{a_{i_1}a_{i_2}\ldots a_{i_k}}$. So, $\S{x}$ is the polynomial that represents all the subsets of $x$. For example, we associate with the itemset $x=abc$ the polynomial $\S{x}=\overline{abc}=(a+1)(b+1)(c+1)$, that gives us the polynomial: $1+a+b+c+ab+ac+bc+abc$.\\
From the itemset subsequence polynomial, we can derive the subsequence polynomial associated with a dataset $D$.
\begin{definition}[Dataset Subsequence Polynomial]
		Let $D = \{t_1,\ldots,t_n\}$ be a dataset. The subsequence polynomial $\S{D}$ associated with $D$ is the sum of the $n$ subsequence polynomials of its transactions:
\begin{equation}
		\S{D} =\sum_{i=1}^{n}\S{t_i}
\end{equation}
\end{definition}
It is obvious to see that the terms of the polynomial $\S{D}$ have the form $\langle\S{D},w\rangle w$, where $w$ is an itemset and $\langle\S{D},w\rangle$ a coefficient in $\N$ representing its support in the database. Indeed, an itemset have 1 as coefficient in the polynomial of the transaction $t_i$ where it appears and, consequently, its coefficient in the database is then the number of the transactions where it occurs. To illustrate this concept let us consider a running example taken from~\cite{Zaki14}. Table~\ref{tab:example1}, shows a database of six transactions, where the third column gives the subsequence polynomial of each transaction. We have calculated also, in the last line, the subsequence polynomial of the whole database. We can easily observe, in the example, that the itemsets: $\epsilon,\,e,\, bc,\, acde$ have the supports : $6, 5, 4$, and $1$ respectively.
\begin{table*}
	\caption{Transaction database and the associated polynomials.}\label{tab:example1} \centering
	\begin{tabular}{|c|l|l|}
		\hline
		$i$ & $t_i$& $ \S{t_i}$\\
		\hline
		1 & abde& $1+a+b+d+e+ab+ad+ae+bd+be+de+abd+ade+abe+bde+abde$\\
		2 & bce&$1+b+c+e+bc+be+ce+bce$\\
		3 & abde& $1+a+b+d+e+ab+ad+ae+bd+be+de+abd+ade+abe+bde+abde$\\
		4 & abce& $1+a+b+c+e+ab+ac+ae+bc+be+ce+abc+ace+abe+bce+abce$\\
		5 & bcd&$1+b+c+d+bc+bd+cd+bcd$\\
		6 & abcde&$1+a+b+c+d+e+ab+ac+ad+ae+bc+bd+be+cd+ce+de+abc+abd$\\
		&&$+abe+acd+ace+ade+bcd+bce+bde+cde+abcd+abce+abde+bcde+acde+abcde$\\
		\hline
		&&$\S{D}=6+4a+6b+4c+4d+5e+4ab+2ac+3ad+4ae+4bc+4bd+5be+2cd+3ce+3de$\\
		&&$+2abc+3abd+4abe+acd+2ace+3ade+2bcd+3bce+3bde+cde$\\
		&&$+abcd+2abce+3abde+bcde+acde+abcde$\\
		\hline
	\end{tabular}
\end{table*}
\subsection{General Algorithm}
\noindent Now, we are given a polynomial $\S{}$ over an alphabet $A$ with coefficients on a semiring $\K$, and a user specified minimum support-threshold $s$. We aim to extract the polynomial $\F$ from $\Sr$ defined as follows: 
\begin{equation}
\langle \F,w \rangle  = 
\begin{cases} \langle \Sr,w \rangle  \qquad \textrm{ if } \; \langle \Sr,w \rangle \ge s,\\
\\
0 \qquad\qquad \textrm{ otherwise.}
\end{cases}
\end{equation}
So, we look for all words from the range of the polynomial $\S{}$ having coefficients greater than $s$. The exploration of the problem space, exponential in nature, is performed by the generic Algorithm~\ref{algo1}, which list the searched set of words by invoking \textsc{Discover-FI}$(\Sr,s,\varepsilon,\emptyset)$. Thanks to the Apriori property in Proposition~\ref{Apriori-prop}, the problem space can be pruned. Note that since the frequentness is a relative notion, we keep in this work, in a similar way as many works~\cite{Cheung03,Goethals04} all the items, regardless of their initial frequencies. This make the model more flexible specially in dynamic datasets.
	
	\begin{proposition}[A-priori~\cite{Agrawal94}]
		Let  $D$ be a dataset and $ w_1,w_2$ two itemsets. If $ w_1 \preccurlyeq w_2$, then   $\sprt{w_1}{D}\ge\sprt{w_2}{D}$.
		\label{Apriori-prop}
	\end{proposition}
	\begin{algorithm}[t]
		\begin{algorithmic}
			\Require The polynomial $\Sr$, the min. support-threshold $s$, and an itemset $w=w_1w_2\ldots w_{|w|}$
			\Ensure The set of all frequent itemsets
			\ForAll  {$a > w_{|w|}$}
			\If {$\langle \Sr,wa \rangle \ge s$}
			\State $\mathcal{F} \gets \mathcal{F} \cup \{(wa, \langle \Sr,wa \rangle)\}$
			\State \textsc{Discover-FI}$(\Sr,s,wa,\mathcal{F}$)
			\EndIf
			\EndFor
		\end{algorithmic}
		\caption{\textsc{Discover-FI}$(\Sr,s,w,\mathcal{F})$}
		\label{algo1}
	\end{algorithm}
	
	It is clear that the complexity of Algorithm~\ref{algo1} depends on the number of FI as well as the cost of the test of frequentness, which depends in turn on the itemset length and the calculation of its coefficient $\langle\Sr,w\rangle$ in the chosen data structure. In order to give efficient implementation of Algorithm~\ref{algo1}, it is necessary to use an optimal data structure, which must have a reduced size and provides a minimal cost of coefficient calculation. In this work, we claim that the FI mining problem can be formulated using formal series which we realize by means of weighted automata~\cite{salomaa78}.
	\section{\uppercase{Frequent Itemset Weighted Automaton}} 
	\noindent  Let $\AS{D}$ be a weighted automaton recognizing the subsequence polynomial $\S{D}$ associated with a dataset $D$ as defined above. Calculate the coefficient $\langle\S{D},w\rangle$ of an itemset $w$ in this polynomial is equivalent to determine its weight in the automaton $\AS{D}$. Consequently, the complexity of this calculation relies on the type of the automaton (deterministic, non-deterministic, asynchronous...etc) and its size. Hereafter, we propose a particular and reduced automaton w.r.t the size of the dataset $D$ which realizes the polynomial $\S{D}$.\\
	For the purpose of the construction of the automaton $\AS{D}$, which we refer as FIWA for Frequent Itemset Weighted Automaton, and since the idea of overlapping common prefixes (prefix tree, trie, prefix relation or equivalence class, FPTree) has proven to be very effective in this problem~\cite{Zaki00,Cheung03,Han04,Valtchev08,Totad12}, we shall go through another type of automaton, which will help us to define our intended automaton $\AS{D}$. This intermediate automaton is the prefixial weighted automaton $\AP{D}$ defined hereafter. But let us define, first, the prefixial polynomial.     
	\begin{definition} [Itemset Prefixial Polynomial]
		Let $x=a_{i_1}a_{i_2}\ldots a_{i_k}$ be a $k$-itemset, the prefixial polynomial associated with $x$ is $\P{x}$ defined as follows: 
		\begin{equation}
		\P{x}=\sum_{u \in \pref{x}}u
		\end{equation}
		
	\end{definition}
	That is, the prefixial polynomial is the sum of all the prefixes of the considered itemset. For example, the prefixial polynomial of the itemset $abc$ is $\P{abc} = 1+a+ab+abc$.
	\begin{definition} [Dataset Prefixial Polynomial]
		Let $D = \{t_1,\ldots,t_n\}$ be a dataset. The prefixial polynomial $\P{D}$ associated with $D$ is the sum of the $n$ prefixial polynomials of its transactions:
		
		\begin{equation}
		\P{D} = \sum_{i = 1}^{n}\P{t_i}
		\end{equation}
	\end{definition}
	Notice that the last definition induces that $\spr{\P{D}} = \pref{D}$. In other words, the range of the prefixial polynomial of a dataset $D$ is the set of the prefixes of its transactions. Below is the prefixial polynomial of the dataset of our running example, after some development: $\P{D} =6+4a+2b+4ab+2bc+2abc+2abd+bcd+bce+abcd+abce+2abde+abcde$.
	
	\subsection{Prefixial Weighted Automaton}
		\noindent At this level, we claim that the construction of a weighted automaton for the dataset subsequence polynomial $\S{D}$ go through the construction of a weighted automaton for $\P{D}$ the prefixial one. There exist many weighted automata that realize these polynomials. We give here, a particular deterministic weighted automaton which realizes the prefixial polynomial $\P{D}$, then introduce a little change on it to get an automaton that realizes our initial dataset subsequence polynomial $\S{D}$.
	\begin{definition}[Prefixial Weighted Automaton (PWA)]
		\label{def9}
		Let $\P{D}$ be the prefixial polynomial of a dataset $D$. The related prefixial weighted automaton $\AP{D} =(Q,A,\mu,\lambda,\gamma)$ is defined as follows:
		\begin{itemize}
			\item $Q = \spr{\P{D}}$, 
			\item $\mu(u)=  1 \text{, for } u=\varepsilon  \text{, and   }  0  \text{ otherwise}$, for $u\in Q$,
			\item $\lambda(u,a,ua) = 1    \text{, for } u \text{ and } ua \in Q \text{, and } a\in A$, 
			\item  $\gamma(u)=  \langle \P{D},u \rangle   \text{, for }  u \in Q$. 
		\end{itemize}
	\end{definition}
	Note that the weight of any path labelled $u$ in a prefixial weighted automaton $\AP{D}$ is equal to $\gamma(u)$, since  $\mu(\varepsilon) = 1$, and $\lambda(v,a,va) = 1$ for all $v,va \in Q$. In in order to alleviate the reading, an automaton $\mathcal{A}$ that realizes $\P{D}$ is said, next, to be PWA if and only if it is isomorphic to $\AP{D}$, i.e: $(\mathcal{A} \cong \AP{D})$. An automaton isomorphic to the prefixial weighted automaton associated with the dataset of our running example is displayed in Figure~\ref{fig1}.      
	\begin{figure*}[ht]
		\vspace{-0.2cm}
		\centering 
			\begin{tikzpicture}[->,>=stealth', shorten >=1pt,auto,node distance=3.5cm,inner sep=0cm,
			semithick,  every state/.style =  {scale = 0.6 , draw,accepting by arrow,font=\footnotesize }, every edge/.style={font=\footnotesize,draw} ]
			\node[initial,initial text ={},state, accepting below, accepting text =6 ] (0) {$0$};
			\node[state, accepting below, accepting text = 4]         (1) [above right of=0] {$1$};
			\node[state, accepting below, accepting text = 4]         (2) [above right of=1] {$2$};
			\node[state, accepting below, accepting text = 2]         (3) [above right of=2] {$3$};
			\node[state, accepting below, accepting text = 1]         (4) [below right of=3] {$4$};
			\node[state, accepting right, accepting text = 1]         (5) [right of=4] {$5$};
			\node[state, accepting right, accepting text = 1]         (6) [ right of=3] {$6$};
			\node[state, accepting below, accepting text = 2]         (7) [below right of=2] {$7$};
			\node[state, accepting right, accepting text = 2]         (8) [right of=7] {$8$};
			\node[state, accepting below, accepting text = 2]         (9) [below right of=0] {$9$};
			\node[state, accepting below, accepting text = 2]         (10) [ right of=9] {$10$};
			\node[state, accepting right, accepting text = 1]         (11) [ above right of=10] {$11$};
			\node[state, accepting right, accepting text = 1]         (12) [ right of=10] {$12$};
			
			\path 
			(0) edge [left]            node[above left] {$a/1$} (1)
			(1) edge [left]             node[above left] {$b/1$} (2)
			(2) edge [left]             node[above left] {$c/1$} (3)
			(3) edge [left]             node[above right ] {$d/1$} (4)
			(4) edge [left]             node[above ] {$e/1$} (5)
			(3) edge [left]             node[above ] {$e/1$} (6)
			(2) edge [left]             node[above right] {$d/1$} (7)
			(7) edge [left]             node[below] {$e/1$} (8)
			(0) edge [left]             node[above right] {$b/1$} (9)
			(9) edge [left]             node[above] {$c/1$} (10)
			(10) edge [left]             node[above left ] {$d/1$} (11)
			(10) edge [left]             node[above] {$e/1$} (12)
			;
			\end{tikzpicture}
			\caption{\footnotesize A PWA associated with our running example dataset.}
			\label{fig1}
		
	\end{figure*}		 
	\begin{lemma}
		For a dataset $D$, the automaton $\AP{D}$ realizes the polynomial $\P{D}$.
	\end{lemma}
	\begin{proof}
		By construction. It is not hard to notice that the boolean automaton derived from $\AP{D}$ (the later deprived from its weights) recognizes the range of the prefixial polynomial $\P{D}$. Indeed, $\AP{D}$ have only one initial state $\varepsilon$ and all the states are final and associated with words in the range of $\P{D}$. Moreover, a transition, if it exists, from a state $u$ is made by items of $A$ leading to $ua$,  which yet remains a word in  the range of $\P{D}$. Furthermore, The weight in the automaton of each word in  the range of $\P{D}$ is exactly its corresponding coefficient, since $\gamma(u) = \langle \P{D},u \rangle$.
	\end{proof}	 
	Definition~\ref{def9}, introduces the prefixial weighted automaton of a dataset from its associated prefixial polynomial. In what follows, we give a construction procedure of this automaton, which can be done in batch or step by step either taking into account one transaction or a set of them.  This process is a general incremental algorithm for the construction of a PWA associated with a dataset $D$.  
	\begin {proposition}
	\label{prop2}
	Let $\mathcal{A}$ and $\mathcal{B}$ be two PWAs associated respectively with datasets $X$ and $Y$. There exists a PWA $\mathcal{C}$ for the dataset $X\cup Y$ derived from  $\mathcal{A}$ and $\mathcal{B}$. \end{proposition}  
\begin{proof}
	The idea is to construct the automaton $\mathcal{C}$ by determinizing both automata $\mathcal{A}$ and $\mathcal{B}$ using the accessible subset-construction procedure. We give below the definition of: $R$ the set of states of the automaton $\mathcal{C}$, and  $\gamma$ the function of final weights (the functions $\mu$ and $\lambda$ are obvious, and remain unchanged as seen in Definition~\ref{def9} and depicted in Figure~\ref{fig2}), and a mapping $h$ from $R$ to the set of states of $\AP{X\cup Y}$, which is the range of the polynomial $\P{X\cup Y}$.\\
	Let $\mathcal{A}=(P,A_1,\mu_1,\lambda_1,\gamma_1)$ be a PWA isomorphic, via $h_1$, to the automaton $\AP{X}$, and  $\mathcal{B}=(Q,A_2,\mu_2,\lambda_2,\gamma_2)$ the one isomorphic, via $h_2$, to the automaton $\AP{Y}$. We define the PWA $\mathcal{C}=(R,A_1\cup A_2,\mu,\lambda,\gamma)$  as follows ($p \in P$ and $q \in Q$):
	\begin{itemize}
		\item Set of states : 
		$ R = \{ \{p,q\} \mid h_1(p) = h_2(q)\}\cup \{\{p\} \mid h_1(p) \in h_1(P)\setminus h_2(Q)\} \cup \{\{q\} \mid h_2(q) \in h_2(Q)\setminus h_1(P)\}$,
		\item Final weights : $\gamma(\{p,q\})=\gamma_1(p)+\gamma_2(q);	\gamma(\{p\})=\gamma_1(p); \gamma(\{q\})=\gamma_2(q)$.  
	\end{itemize}
	The mapping $h$ from $R$ to $\spr{\P{X\cup Y}}$  as follows :\\
	$h(\{p,q\}) =h_1(p);\quad h(\{p\}) = h_1(p);\quad h(\{q\}) = h_2(q)$.\\
	There is no difficulty to verify that the mapping $h$, as defined above, is a weighted automata isomorphism which is omitted here for space limitation.
\end{proof}	
In Figure~\ref{fig2},  we illustrate the above construction by an example of merging and determinizing of two simple PWA associated with the following two datasets $X=\{ab,ac\},Y=\{ac,ad,e\}$
\begin{figure*}[ht]
	\centering
	\begin{minipage}[bl]{.4\linewidth}
		\subfigure[]{
		\begin{tikzpicture}[->,>=stealth', shorten >=1pt,auto,node distance=3cm,inner sep=0cm,
		semithick,  every state/.style =  {scale = 0.7 , draw,accepting by arrow,font=\footnotesize }, every edge/.style={font=\footnotesize,draw} ]
		\node[initial,initial text ={},state, accepting below, accepting text =2 ] (0) {$p_0$};
		\node[state, accepting below, accepting text = 2]         (1) [right of=0] {$p_1$};
		\node[state, accepting below, accepting text = 1]         (2) [right of=1] {$p_2$};
		\node[state, accepting right, accepting text = 1]         (3) [ above   of=1] {$p_3$};
		
		\path 
		(0) edge [left]            node[below] {$a/1$} (1)
		(1) edge [left]             node[below] {$b/1$} (2)
		(1) edge [left]             node[left] {$c/1$} (3);
		\end{tikzpicture}}
		
	\end{minipage} \hspace*{1.5cm}
	\begin{minipage}[br]{.4\linewidth}
		\subfigure[]{
		
		\begin{tikzpicture}[->,>=stealth', shorten >=1pt,auto,node distance=3cm,inner sep=0cm,
		semithick,  every state/.style =  {scale = 0.7 , draw,accepting by arrow,font=\footnotesize }, every edge/.style={font=\footnotesize,draw} ]
		\node[initial,initial text ={},state, accepting below, accepting text = 3] (0) {$q_0$};
		\node[state, accepting below, accepting text = 2]         (1) [right of=0] {$q_1$ };
		\node[state, accepting below, accepting text = 1]         (2) [right of=1] {$q_2$ };
		\node[state, accepting right, accepting text = 1]         (3) [above  of=0] {$q_4$};
		\node[state, accepting right, accepting text = 1]         (4) [above  of=1] {$q_3$};
		\path 
		(0) edge [left]            node[below] {$a/1$} (1)
		(0) edge [left]            node[left] {$e/1$} (3)
		(1) edge [left]              node[below] {$d/1$} (2)
		(1) edge [left]             node[left] {$c/1$} (4);
		\end{tikzpicture}}
	\end{minipage}
  
	\centering
		\begin{minipage}[b]{.4\linewidth}
			\subfigure[]{
			
			\begin{tikzpicture}[->,>=stealth', shorten >=1pt,auto,node distance=3cm,inner sep=0cm,
			semithick,  every state/.style =  {scale = 0.8 , draw,accepting by arrow,font=\footnotesize }, every edge/.style={font=\footnotesize,draw} ]
			\node[initial,initial text ={},state, accepting below, accepting text =5 ] (0) {$\{p_0,q_0\}$};
			\node[state, accepting below, accepting text = 4]         (1) [right of=0] {$\{p_1,q_1\}$};
			\node[state, accepting below, accepting text = 1]         (2) [right of=1] {$\{p_2\}$};
			\node[state, accepting right, accepting text = 1]         (3) [ above  of=0] {$\{q_4\}$};
			\node[state, accepting right, accepting text = 1]         (4) [ above right  of=1] {$\{q_2\}$};
			\node[state, accepting right, accepting text = 2]         (5) [ above of=1] {$\{p_3,q_3\}$};
			\path 
			(0) edge [left]            node[below] {$a/1$} (1)
			(0) edge [left]            node[left] {$e/1$} (3)
			(1) edge [left]             node[below] {$b/1$} (2)
			
			(1) edge [left]             node[left] {$c/1$} (5)
			(1) edge [left]             node[below right ] {$d/1$} (4);
			
			\end{tikzpicture}}
	
		\end{minipage}

	\caption{\footnotesize two automata (a) and (b) and the merging one (c) by determinization.}	
	\label{fig2}
\end{figure*}

\subsection{Analysis of the Prefixial Weighted Automata Merging Construction}
\noindent The previous procedure, in Proposition~\ref{prop2}, introduces a construction method of a PWA associated with a dataset. More interesting, it makes no assumptions about the fragments $X$ and $Y$, and therefore, it provides a flexible construction algorithm of the union of two or more PWAs, either in batch or incremental way.\\
Moreover, this construction offers some complexity-related remarkable properties. Here, we mention some of them.\\
The following lemma is induced from the inclusion-exclusion principle.
\begin{lemma} 
	Let X and Y be two datasets. Then:
	\begin{equation}
	|\AP{X \cup Y}| \le |\AP{X}|+|\AP{Y}|
	\end{equation}
\end{lemma} 
Consequently, we obtain this two corollaries about the size of a PWA and the complexity of its construction.  
\begin{corollary}
	Let $D$ be a dataset. Then:
	
	\begin{equation}
	|\AP{D}| \le |D| 
	\end{equation}
\end{corollary}
Likewise, and as the subset construction of a PWA automaton associated with the dataset $X\cup Y$ derived from the PWAs associated with $X$ and $Y$ is guided by the transitions of the smallest automaton, we can state the following lemma. 
\begin{lemma}
	A PWA associated with the dataset $X \cup Y$ can be constructed from the PWAs associated with $X$ and $Y$ in $O(\min(|\AP{X}| ,|\AP{Y}|))$ time.
\end{lemma}
Consequently, we can deduce the following Proposition.
\begin{proposition}
	Let $D$ be a dataset. A PWA of $D$ can be constructed in $O(|D|)$ time and space complexity.	
\end{proposition}
Further, we can naturally generalize these results to $k$ datasets. This constitutes an important criterion, that provides a fluid tuning and a flexible data partitioning scheme, which is very useful in many aspects of the problem. Indeed, it allows to deal with the memory requirements, parallelization and/or incrementality constraints, since, it does not matter, here, the granularity of this partitioning: by transaction as in~\cite{Cheung03}, or by batch as in~\cite{Totad12}, taking two or more data fragments. The following corollary gives this extension.
\begin{corollary}
	Let $\mathcal{A}_1,\ldots, \mathcal{A}_k$ be $k$ PWA associated respectively with the datasets $X_1,\ldots,X_k$ . One can construct a PWA associated with the union $X_1\cup\ldots\cup X_k$ in $O(|X_1|+\ldots+|X_k|)$ time and space complexity.
	\label{Cor13}
\end{corollary}
\subsection{Toward the Itemset Weighted Automaton}
	\noindent The work done, so far, is a significant step toward our objective. Recall that our goal is to construct a weighted automaton that realizes the subsequence polynomial $\S{D}$ associated with a dataset $D$. Let us define here another polynomial, which we refer to as the prefixial-bar polynomial. The later serves as an intermediate one, that guides us to obtain the targeted one $\S{D}$.
\begin{definition}
	\label{def10}
	Let $D$ be a dataset, and $\P{D}$ the associated prefixial polynomial. The prefixial-bar polynomial $\overline{\P{D}}$ is:
	
	\begin{equation}
	\overline{\P{D}} =  \langle \P{D}, \varepsilon \rangle +\sum_{ \stackrel{u \in A^*}{a \in A}} \langle \P{D}, ua \rangle \overline{u}a
	\end{equation}

\end{definition} 
Obviously, the prefixial-bar polynomial of the dataset $D$ depends on the prefixial one. We give the following proposition.
\begin{proposition}
	\label{prop3}
	Let $D=\{t_1,\ldots,t_n\}$ be a dataset. Let $\b{\P{D}}$ and $\S{D}$ respectively the associated prefixial-bar and the subsequence polynomials. Then:
	\begin{equation}
	\overline{\P{D}} = \S{D}
	\end{equation}
	
\end{proposition}

\begin{proof}
	Let us start by checking that the Proposition~\ref{prop3} is true for one transaction $t_i$ taken from the dataset $D$ of $n$ transactions.\\
	So, let $t_i=a_{i_1}a_{i_2}\ldots a_{i_k}$ be a $k$-itemset. According to the definitions in Sections 3 and 4, and the convention $\overline{a_i}=a_i+1$, we have:
	\begin{eqnarray*}
		\P{t_i}&=& 1+a_{i_1}+a_{i_1}a_{i_2}+\ldots+a_{i_1}a_{i_2}a_{i_3} \ldots a_{i_k}\\
	\text{So, }\b{\P{t_i}}&=&1+a_{i_1}+\overline{a_{i_1}}a_{i_2}+\ldots+\overline{a_{i_1}a_{i_2} \ldots a_{i_{k-1}}}a_{i_k}\\
		&=&\overline{a_{i_1}}+\overline{a_{i_1}}a_{i_2}+\ldots+\overline{a_{i_1}a_{i_2}a_{i_3} \ldots a_{i_{k-1}}}a_{i_k} \\
		&=&\overline{a_{i_1}a_{i_2}}+\ldots+\overline{a_{i_1}a_{i_2}a_{i_3} \ldots a_{i_{k-1}}}a_{i_k} \\
		&\ldots&\\
		&=& \overline{{a_{i_1}a_{i_2}a_{i_3} \ldots a_{i_{k-1}}a_{i_k}}}\\
		&=&\S{t_i}	
	\end{eqnarray*}
	Now let us verify also the equality between the sum of the prefixial-bar polynomials and the prefixial-bar polynomial of the whole dataset $D$.
	\begin{eqnarray*}
		\b{\P{t_i}} &=&  \langle \P{t_i}, \varepsilon \rangle + \sum_{ \stackrel{u \in A^*}{a \in A}} \langle \P{t_i}, ua \rangle \overline{u}a\\
		\qquad \sum_{i = 1}^{n}\overline{\P{t_i}} &=& \sum_{i = 1}^{n} \langle \P{t_i}, \varepsilon \rangle +\sum_{i=1}^{n} \sum_{\stackrel{u \in A^*}{a \in A}}\langle \P{t_i}, ua \rangle \overline{u}a\\
		&=& \sum_{i = 1}^{n} \langle \P{t_i}, \varepsilon \rangle +\sum_{\stackrel{u \in A^*}{a \in A}} \sum_{i=1}^{n} \langle \P{t_i}, ua \rangle \overline{u}a\\	
		&=& \langle \P{D}, \varepsilon \rangle +\sum_{\stackrel{u \in A^*}{a \in A}} \langle \P{D}, ua \rangle \overline{u}a\\
		&=&\overline{\P{D}}
	\end{eqnarray*}
	We've found that:
	 $\overline{\P{t_i}} = \S{t_i}
	  \text{, so }\displaystyle\sum_{i=1}^{n}\overline{\P{t_i}} = \sum_{i = 1}^{n}\S{t_i}\;\\
	 \text{which leads to } \; \overline{\P{D}} = \S{D}$. 
\end{proof}
The construction of an automaton that compute the dataset subsequence polynomial $\S{D}$ become now easier. Note that the polynomial $\b{\P{D}}$ can be rewritten to show the link with the polynomial $\P{D}$ by adding null terms:
$$
\overline{\P{D}} =  \langle \P{D}, \varepsilon \rangle +\sum_{ u \in\, \spr{\P{D}}}0\times\b{u}+ \sum_{ \stackrel{ua \in\, \spr{\P{D}}}{a\in A} } \langle \P{D}, ua \rangle \overline{u}a
$$
By bringing together the expressions of $\P{D}$ and that of $\b{\P{D}}$, we can note the bijection between each $u$ in $\P{D}$ and $\b{u}$ in $\b{\P{D}}$. Consequently, since $\b{u}$ encodes the subsequences of $u$ (see Definition~\ref{def1}), it suffices, thus, to add $\varepsilon$-transitions in paths labelled $u$ in our automaton $\AP{D}$; However, we must be scrupulous about coefficients, because adding $\varepsilon$-transitions may multiply the recognition paths of an itemset. This can be fixed by state duplication. That is, for each state $u\ne\varepsilon$, we create a second one ($\b{u}a$) with the right coefficient ($\langle \P{D}, ua \rangle$), the original becomes a non-accepting state with null weight ($0\times \b{u}$). Notice that this state/transition duplication is, here, artificial and will be simulated as shown in Algorithm~\ref{algo3}. This trick also insures the values of the other terms in the rest of the polynomial $\b{\P{D}}$. We illustrate this idea by a simple example of a dataset $D$ containing only two transactions $D= \{ abc,ab\}$. In Figure~\ref{fig3}, we give the two automata: a PWA of $D$, and the extended one.

\begin{figure*}[ht]
	\vspace{-0.2cm}
	\centering
	\begin{minipage}[l]{.8\linewidth}
		\centering 
		\subfigure[]{
		
		\begin{tikzpicture}[->,>=stealth', shorten >=1pt,auto,node distance=3.5cm,inner sep=0cm,
		semithick,  every state/.style =  {scale = 0.65 , draw,accepting by arrow,font=\footnotesize }, every edge/.style={font=\footnotesize,draw} ]
		\node[initial,initial text ={},state, accepting above, accepting text =2 ] (0) {$\varepsilon$};
		\node[state, accepting above, accepting text = 2]         (1) [right of=0] {$a$};
		\node[state, accepting above, accepting text = 2]         (2) [right of=1] {$ab$};
		\node[state, accepting above, accepting text = 1]         (3) [right of=2] {$abc$};
		\path 
		(0) edge [left]            node[above] {$a/1$} (1)
		(1) edge [left]             node[above] {$b/1$} (2)
		(2) edge [left]             node[above] {$c/1$} (3);
		\end{tikzpicture}}

	\end{minipage} \vspace{0.1cm}
	
	\begin{minipage}[c]{.8\linewidth}

	\centering	
	\subfigure[]{
		
		\begin{tikzpicture}[->,>=stealth', shorten >=1pt,auto,node distance=3.5cm,inner sep=0cm,
		semithick,  every state/.style =  {scale = 0.65 , draw,accepting by arrow,font=\footnotesize }, every edge/.style={font=\footnotesize,draw} ]
		\node[initial,initial text ={},state, accepting below, accepting text = 2] (0) {$\b{\varepsilon}$};
		\node[state, accepting below, accepting text = 0]         (1) [right of=0] {$\b{a}$ };
		\node[state, accepting below, accepting text = 0]         (2) [right of=1] {$\b{ab}$};
		\node[state, accepting below, accepting text = 0]         (3) [right of=2] {$\b{abc}$};
		
		\node[state, accepting right, accepting text = 2,draw,densely dotted]      (4) [above right of=0] {$\b{\varepsilon}a$};
		\node[state, accepting right, accepting text = 2,draw,,densely dotted]     (5) [above right of=1] {$\b{a}b$};
		\node[state, accepting right, accepting text = 1,draw,,densely dotted]     (6) [above right of=2] {$\b{ab}c$};

		\path 
		(0) edge [left]            node[above] {$a/1$} (1)

		(1) edge [left]              node[above] {$b/1$} (2)				
		(2) edge [left]             node[above] {$c/1$} (3);

		\path[densely dotted]
		(1) edge [left]             node[left] {$b/1$} (5)
		(0) edge [left]             node[left] {$a/1$} (4)
		(2) edge [right]            node[left] {$c/1$} (6)
		(0) edge [bend right]            node[below] {$\varepsilon/1$} (1)
		(1) edge [bend right]            node[below] {$\varepsilon/1$} (2)
		(2) edge [bend right]            node[below] {$\varepsilon/1$} (3);
		\end{tikzpicture}}	
	\end{minipage} 
	\caption{\footnotesize a PWA (a) and its extended automaton (b). }	
	\label{fig3}
\end{figure*}

\subsection{The Mining Algorithm}
	\noindent Once the PWA associated with $D$ has been built using one of the processes introduced by  the Proposition~\ref{prop2} or Corollary~\ref{Cor13}, it serves as a structure for the problem space exploration. In our mining phase, we explore the automaton using a depth-first traversal as exhibited in Algorithm~\ref{algo2}. The main strength of our algorithm is that it doesn't require any additional memory other than that needed for the WPA as opposed to the previous approaches (see~\cite{Goethals04}).  
\begin{algorithm}[ht]
	\begin{algorithmic}
		\Require a PWA of $D$, the support-threshold $s$, a set of states $Q_w$, and an itemset $w$
		\Ensure The set of all FI
		\ForAll  {$a > w_{|w|}$}
		\State $(Q_{wa},\langle \S{D},wa \rangle) \gets \Call {Extend}{Q_w,w,a}$
		\If {$\langle \S{D},wa \rangle \ge s$}
		\State $\mathcal{F} \gets \mathcal{F} \cup \{(wa, \langle \Sr,wa \rangle)\}$
		\State \textsc{Discover-FI}$(\Sr,s,wa,\mathcal{F}$)
		\EndIf
		\EndFor
	\end{algorithmic}
	\caption{\textsc{Discover-FI}$(\Sr,s,w,\mathcal{F})$}
		\label{algo2}	
\end{algorithm}
\begin{algorithm}[ht]

	\begin{algorithmic}
			\Require set of states $Q_w$, an itemset $w$, an item $a$
			\Ensure The extended set of states $Q_{wa}$
		\State  $P \gets  Q_w$
		\State $R \gets \emptyset$ 
		\While {$P \ne \emptyset$}
		\State $q \gets$  pick a state from $P$
		\If{ $i(q) = a$} \State $R \gets R \cup \{q\}$
		\ElsIf{ $i(q)<a$}
		\State $P \gets P \cup \delta^+(q)$
		\EndIf
		\EndWhile
		\State \textbf{return} $(R,\gamma(R))$
	
		\caption{\textsc{Extend}$(Q_w,w,a)$}
		\label{algo3}
	\end{algorithmic}
\end{algorithm}

The exploration begins with the invocation \textsc{Discover-FI}$(\Sr,s,\{q_0\},\varepsilon, \emptyset)$, where $q_0$ is the initial state of the automaton. At each step, and starting from the set of states $Q_w$, an itemset $w$ is extended by concatenation with its successors by calling the function \textsc{Extend}$(Q_w, w, a)$. This call returns the set of states $Q_{wa}$ of all paths labeled $wa$ with their coefficients. The support of the concerned itemset $wa$ is then the sum of the coefficients of the elements in the returned set $Q_{wa}$, since $\gamma(R)=\displaystyle\sum_{r \in R}\gamma(r)$. If this extension succeeds with a frequent itemset, the process will continue taking into account the last reached set of states $Q_{wa}$; Otherwise the returned couple is ($\emptyset,0$). Note that $i(q)$ stands for the item-label of the transition leading to the state $q$, with $i(q_0)=\varepsilon$, and $\delta^+(q)$ for the successor states of the state $q$.

\begin{proposition}
	Algorithm~\ref{algo2} can be done in $O(\sum\limits_{{\stackrel{w\in \mathcal{F}}{a > w_{|w|}}}}C_{wa})$ time and $O(|Q|)$ space, where $Q$ is the set of states of the PWA, and $\mathcal{F}$ is the set of FI in the dataset, and $C_{wa}$ is the time required to compute the set $Q_{wa}$ by extending the set $Q_w$.  
	\label{prop4}
\end{proposition}
\begin{proof}
	Our automaton is acyclic over a sorted alphabet. So, the length of any path is upperbounded by $|Q|$. $C_{wa}$ is the time needed to the call of the function \textsc{Extend}, which computes the set $Q_{wa}$ taking into account the last obtained set of states $Q_w$. Let, without loss of generality, $C_{wa}=|Q_{wa}|$, hence for each itemset $w=w_1\ldots w_k$ in $\mathcal{F}$, we have: $C_{w_1}+C_{w_1w_2}+\ldots+C_{w_1w_2\ldots w_k} <|Q|$. Consequently, if $\mathcal{F}_M$ denote the set of maximal frequent itemsets, and $\mathcal{F}_L$ the set of maximal frequent itemsets w.r.t to the prefixial relation $(\mathcal{F}_M \subseteq \mathcal{F}_L \subseteq \mathcal{F})$, we obtain the inequality : $|\mathcal{F}_M||Q|\le \sum_{w\in\mathcal{F}}C_{wa}\le|\mathcal{F}_L||Q|\le|\mathcal{F}||Q|\le|\mathcal{F}||D|$. Further, the memory requirement of the recursive exploration is also upperbounded by $|Q|$; it does not matter the length of the itemset to be recognized or the current level of the exploration, since the returned sets, during the traversal, are pairwise disjoint and their union is $Q$ in the worst case.    	
\end{proof}
\section{\uppercase{Comparison and Unification}}
\noindent 
A theoretical framework based on formal concept analysis and lattice theory is presented early in~\cite{Godin95,Zaki98}. Recently, in ~\cite{Pijls10} attempt is made to unify the common FI-algorithms w.r.t the traversal paradigms well-known in the operations research community. \\
Our model uses formal series, which are mappings between a monoid $\M$ and a semiring $\K$. The appropriate choice of $\M$ and $\K$, and the automaton characteristics which realizes it is driven by the targeted application and needed performances. For the basic version of the FI-mining problem, that is mining itemsets, we opted for the counting semiring $(\N,+,\times,0,1)$ because it offers an intuitive and easy implementation.\\
We are convinced that this framework can be generalized for mining other elaborated items such as sequences, trees or graphs, provided that much more work must be carried out to define monoids of these elements with the appropriate operations and the corresponding implementations by means of specific automata.\\
In what follows, we compare our model against the main state of the art techniques, and explain how these ones can be derived from it.\\
\textbf{Level-wise Approaches :} An Apriori-like algorithm~\cite{Agrawal94} proceeds level by level. First, it computes the frequent singletons and then forms from these a set of candidate doublets. After determining the frequent doublets, it continues to generate the set of frequent triplets and so on, until no new frequent itemsets can be generated. Despite its limits: generating a huge number of candidates and repetitive database scans, this algorithm stay one of the top cited algorithms in the DM community~\cite{Wu08}. Our model can be modified to fit a similar principle if we use an adapted deterministic version of the defined automaton, and perform a simple linear traversal of it in a stepwise fashion. Notice that this adaptation to Apriori allows to devise a more efficient algorithm, since in one hand any itemset have only a unique acceptation path, and in the other hand we do not make use of candidate generation neither database scans for support computation.\\
\textbf{Vertical Approaches :}
The main benefit of the vertical approach~\cite{Zaki00} against the level-wise approaches is speedy in the support calculation via set intersections. However, the drawback as mentioned in the introduction and by the author itself in an improved version is when the intermediate results become too big. Our method, in contrast, is based on a simple output weight read or their summation without need of any additional memory.\\
The vertical approach can be seen as a formal series on the powerset semiring of the set $T$ of transactions $(2^T,\cup,\cap, \emptyset,T)$, with the $\min$ operation computed by set intersection, and the sum by the union. The weight of a transition represents the cardinality of the tidlist of the itemset formed by the path from the root $\emptyset$ to the considered node.\\
\textbf{Projection Approaches :}
It seems to the first glance that our defined automaton is an FPTree~\cite{Han00} by an other way. We must emphasize at the outset that the similarity to FPTree or other concepts in any of the previous work should be seen as a positive point and not the inverse, since our purpose is the definition of a unifying model. We claim, in the other hand, that this is not correct enough. First of all, our automaton is not a data structure but rather a computational model, which can be implemented in different ways. Secondly, The mining algorithms are significantly different. While FPGrowth use a heavily intermediate memory, and also time overhead, for conditional databases and conditional FPTrees construction, our model do not require any additional memory other that necessary for the automaton. Further, and unlike FPGrowth, the open ordering adopted in our model leads to significant time improvement both in the construction phase (only one scan is required), and the mining one, since there is no need to repetitive resorting, neither database projections. Additionally, we argue that our approach outperforms also extension of FPTrees like CATSTree~\cite{Cheung03}, which the building may require many node swaps to maintain its integrity (the support of a parent must be greater than the sum of its children's supports), and incurs consequently some overhead. To the end of  unification, we can view these approaches as a sequence of right derivations by the set of items $A$ of our dataset subsequence polynomial $\S{D}$, or like the exploration of the mirror of the automaton. Indeed, the right derivative of the polynomial $\S{D}$ w.r.t an item $a$ produces the polynomial representation of the $a$-conditional database in FPGrowth.\\
\textbf{Tropical Semiring :} 
An equivalent modeling to our approach can be obtained by using the tropical semiring, computed by a different weighted automaton, where the transitions carry the output weights. In this case, the output weights of all states are $\infty$. The weight of a path is the minimum of the weights of its transitions. It is obvious to note that this model, although equivalent, is expensive compared to which we have adopted, that consists to a simple read of the state output weight.  
\section{\uppercase{Conclusion}}
\noindent 
We have proposed a new model for mining FI.  This model is based on formal series over the semiring $(\N,+,\times,0,1)$, whose the range constitutes the itemsets and the coefficients their supports. We argue that the strength of the introduced formalism are numerous. First, while remaining simple and intuitive, it is complete to model the basic problem. Secondly, it allows the decomposition of the problem to deal, eventually, with the constraints of time or space or both, into independent sub-problems which leads to parallelization and/or incrementatlization processes. Furthermore, the proposed model can be generalized to handle more complex items such as sequences, trees...etc. On the practical side, the model provides an implementation whose performance are proved to be competitive.\\
We have also, reduced this problem, in its basic version, to that of word recognition, allowing an implementation without extra memory in $O(|\mathcal{F}_L||Q|)$ time and $O(|Q|)$ space.\\
In future work, we can improve the mining algorithm by avoiding to recompute the extensions for itemsets $u$ and $v$ having the same returned set of states after the call to the function \textsc{Extend} ($Q_u=Q_v$). This can be done by working on the deterministic automaton equivalent to $\AS{D}$. We will show, in a subsequent work, that this optimization gives also a new time upperbound, which is the number of states of this deterministic automaton, which we conjecture will not be exponential. Furthermore, despite that it is not trivial, it would be very interesting to study other properties of the defined automaton such as its minimization.\\
We also plan to extend this approach to mine, first,  the set of frequent maximal and closed itemsets, and then sequences and trees. Finally, the algebraic aspects of formal series deserves more investigation, and might lead to other theoretical or practical results.     
\vfill
\bibliographystyle{apalike}
{\small
	\bibliography{sample}}
	
\end{document}